\documentclass[twocolumn,prl,aps,epsf,showpacs]{revtex4}

\usepackage{graphicx}
\usepackage{dcolumn}
\usepackage{bm}

\begin{document}


\title{Flow Conductance of a Single Nanohole}

\author{M. Savard$^{1}$, C. Tremblay-Darveau$^{1}$ and G. Gervais$^{1\ast}$ }

\affiliation{$^{1}$Department of Physics, McGill University,
Montreal, H3A 2T8, CANADA}
\date{\today }

\begin{abstract}

The mass flow conductance of single nanoholes with diameter ranging from 75 to 100 nm was measured using mass spectrometry. For all nanoholes, a smooth crossover is observed between single-particle statistical flow (effusion) and the collective viscous flow emanating from the formation of a continuum. This crossover is shown to occur when the gas mean free path matches the size of the nanohole diameter. As a consequence of the pinhole geometry, the breakdown of the Poiseuille approximation is observed in the power-law temperature exponent of the measured conductance.

\end{abstract}

 \maketitle

The concept of resistance to flow, or conversely the flow conductance $G$, dates back to the early development of fluid mechanics. Given a transport channel with a specific geometry connected to two reservoirs, a {\it source} ($S$) and {\it drain} ($D$), a mass flow current $Q=G\cdot \Delta P$ is induced whenever a pressure difference $\Delta P$ is provided across the channel.  This pressure-mass current relationship is analogous to the electrical current flowing  in a resistor with a resistance $R=G^{-1}$ when a voltage drop is supplied across it, $I=G\cdot \Delta V$. Despite fundamental differences in the nature of the constituents forming the reservoir, {\it i.e.} electrons carry a charge whereas particles in a fluidic system are usually neutral, much of the transport processes are very alike and the problem often reduces to the measurement and/or the calculation of the conductance of the channel. Here, we report on the first measurement of the gas flow conductance of a {\it single hole} of nanometric size.  A direct observation of {\it Knudsen effusion} \cite{GRUENER08}
is made whose crossover transition is found to occur when the mean free path of the gas matches the nanohole diameter, as expected. 


{\it A priori} the concept of conductance remains well-defined for transport channels as large as a macroscopic conductor or pipe, or as small as a nano-fabricated channel. With the recent progress made in microfluidics \cite{STONE04} and the push towards nanofluidics \cite{SCHOCH08} for molecular detection \cite{HUH07,FU07}, experimental investigation of the transport properties of fluids at the nanoscale becomes more accessible. With the exception of a few experiments involving quantum fluids \cite{AVENEL85,PACKARD97,Davis,PEARSON01} and DNA sensing \cite{DEKKER07}, most experimental investigations have so far been restricted to either a single pipe in the micron range \cite{EWART07}, or to porous membranes with sizes and properties averaged over the very large number of nanostructures ({\it e.g}  $\sim 10^{11}$ nanochannels/cm$^{2}$) \cite{HOLT06,WHITBY07,GRUENER08}. The properties of flow, hence the conductance, for a single nano-fabricated hole has been left out between these two limits, not by lack of interest \cite{SHARIPOV98,ROY03}, but rather by the difficulties of measuring the very small flow emanating from a single nanochannel. 


 Figure 1A shows a realistic drawing of the experimental cell used for the gas flow measurements and Fig. 1B shows a field-emission transmission electron microscope (FE-TEM) image of the nanohole with a circle of diameter $101\pm 2$ nm which best fits the aperture. The FE-TEM was also used to drill the hole through a $50$ nm thick window of dimension $30\times 40$ micrometer, made of low-stress Silicon Nitride (SiN).  The wafer was  
 then epoxy-sealed on a sample holder (see Fig. 1A middle piece) that was subsequently  inserted into the body of the experimental cell.  Both the lid (top part of cell in Fig. 1A) and the sample holder were sealed with soft indium o-rings to protect against leaks to the outside and leaks around the holder, respectively. The drain pressure below the membrane ($P_{D}$) is kept at vacuum through continuous pumping and helium gas is introduced in the top part of the cell creating a pressure gradient $\Delta P = P_{S}-P_{D}\simeq P_{S}$ which  induces a mass flow $Q$. This flow was detected with a Pfeiffer vacuum Smart Test HLT560 calibrated with an external standard leak of $2.79\times10^{-8}$ atm$\cdotp$ cc/s $\pm 10-15$\%. A cartoon representation of the whole experiment is shown in Fig. 1C. The two reservoirs are depicted by a capillary conductance $G_{S}$ and $G_{D}$, in series before and after the nanohole with a conductance $G_{nh}$.  The mass spectrometer is denoted by $A_{M}$ and measures the total mass current $Q$ when the drain side of the set-up is kept under a vacuum, typically below $\sim 2\cdotp{10^{-3}}$ mbar. With our technique, the total conductance $G_{T}^{-1}=G_{S}^{-1}+G_{D}^{-1}+G_{nh}^{-1}$ of the circuit  is measured. The source and drain conductance can be estimated using the infinite pipe approximation for Poiseuille flow ($G_S\sim  10^{-11}$ m$\cdot$s at $\sim 1$ bar) and Knudsen free-molecular diffusion ($G_{D}\sim  10^{-13}$ m$\cdot$s  at $10^{-3}$ mbar). These conductances are several orders of magnitude larger than the nanohole conductance which has a typical value $G_{nh} \sim 10^{-18}$ m$\cdot$s (see Fig. 2B).  We therefore can approximate $G_{T}\simeq G_{nh}$ in our measurement circuit to a very good accuracy. 

\begin{figure}[tbp]
\includegraphics[width=0.9\linewidth,angle=0,clip]{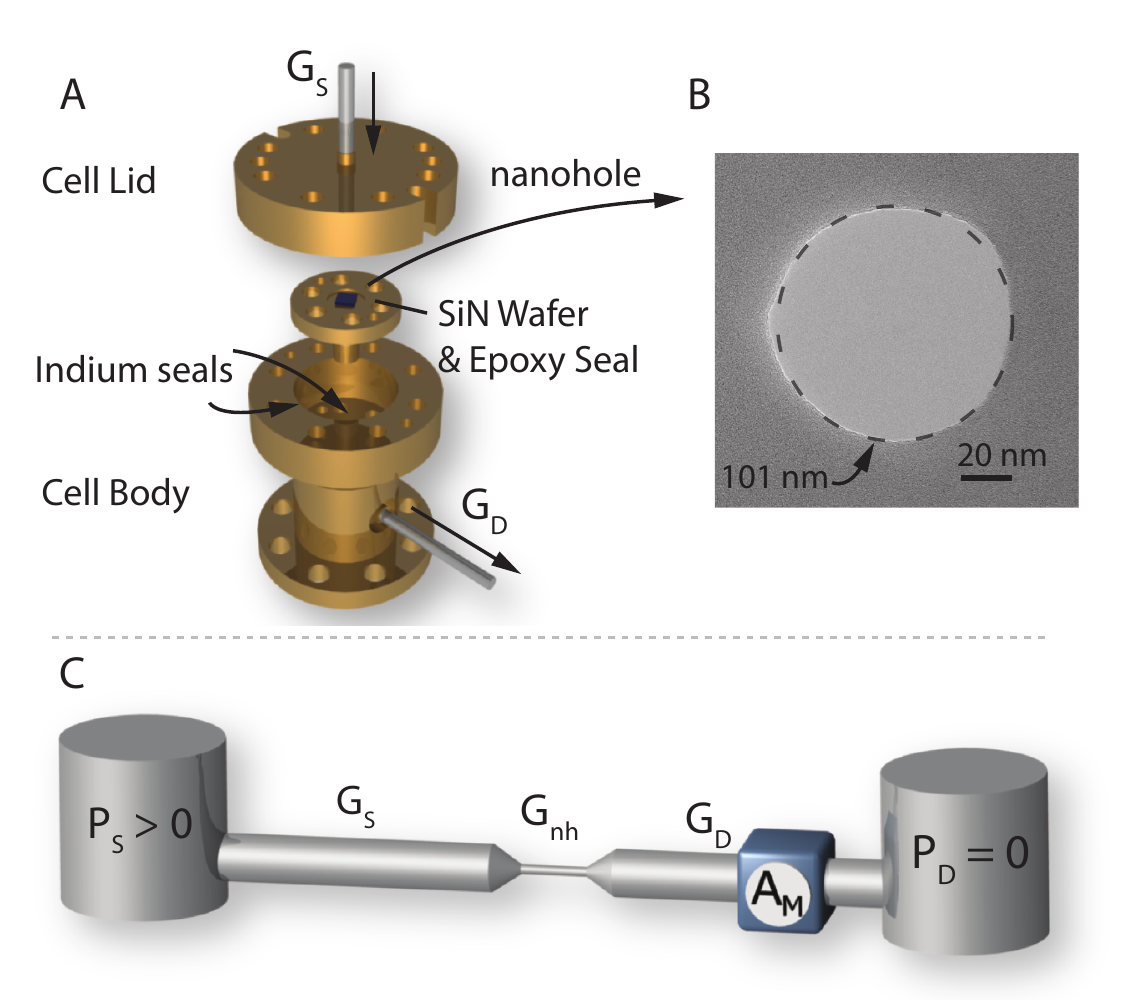}
\caption{\textbf{A}, Experimental cell for the mass flow conductance experiment.  Gas inlet and outlet are connected through stainless steel capillaries with conductance $G_S$ and $G_D$. \textbf{B}, TEM image of the nanopore with diameter $101\pm 2$ nm. \textbf{C}, Cartoon  of the experimental set-up. A gas reservoir held at pressure $P_S$ induces a flow  through the nanohole with a conductance $G_{nh}$ (not drawn to scale), whereas a second reservoir is kept under vacuum, $P_{D}=0$. The mass current flow is measured with a mass spectrometer analyzer $A_{M}$. 
}
\label{figure1}
\end{figure}

Typical measurements of the mass flow through the nanohole are shown in Fig. 2A for pressures up to $38$ bar and at temperature $T=77$ K. These data are the raw output of the mass spectrometer which measures flows in units of [mbar$\cdot$l/s], and  plotted versus the source pressure $P_{S}$. The uncertainty in the calibration of the spectrometer induces a small offset in the measured flow. To correct for this instrumental artifact, a constant was added to the measured mass flow so that the condition $Q\rightarrow 0$ with $P_{S}\rightarrow 0$ was enforced. Three distinct pressure ranges are shown to emphasize the linearity (shown with a dotted line) of the mass flow  with pressures in the range  $P_{S}\simeq 0-55$ mbar, and the strong and weak departure from linearity for pressures $P_{S}\sim 0.69-5$ bar and $P_{S}> 6.2$ bar. The non-linear departure of the flow with applied pressure is the first indication of a non-constant conductance, {\it i.e.}  evolving with  a functional dependence $G=G(P_{S})$ in the pressure range $P_{S}=0\sim 38$ bar.\\

\begin{figure}[tbp]
\includegraphics[width=1\linewidth,angle=0,clip]{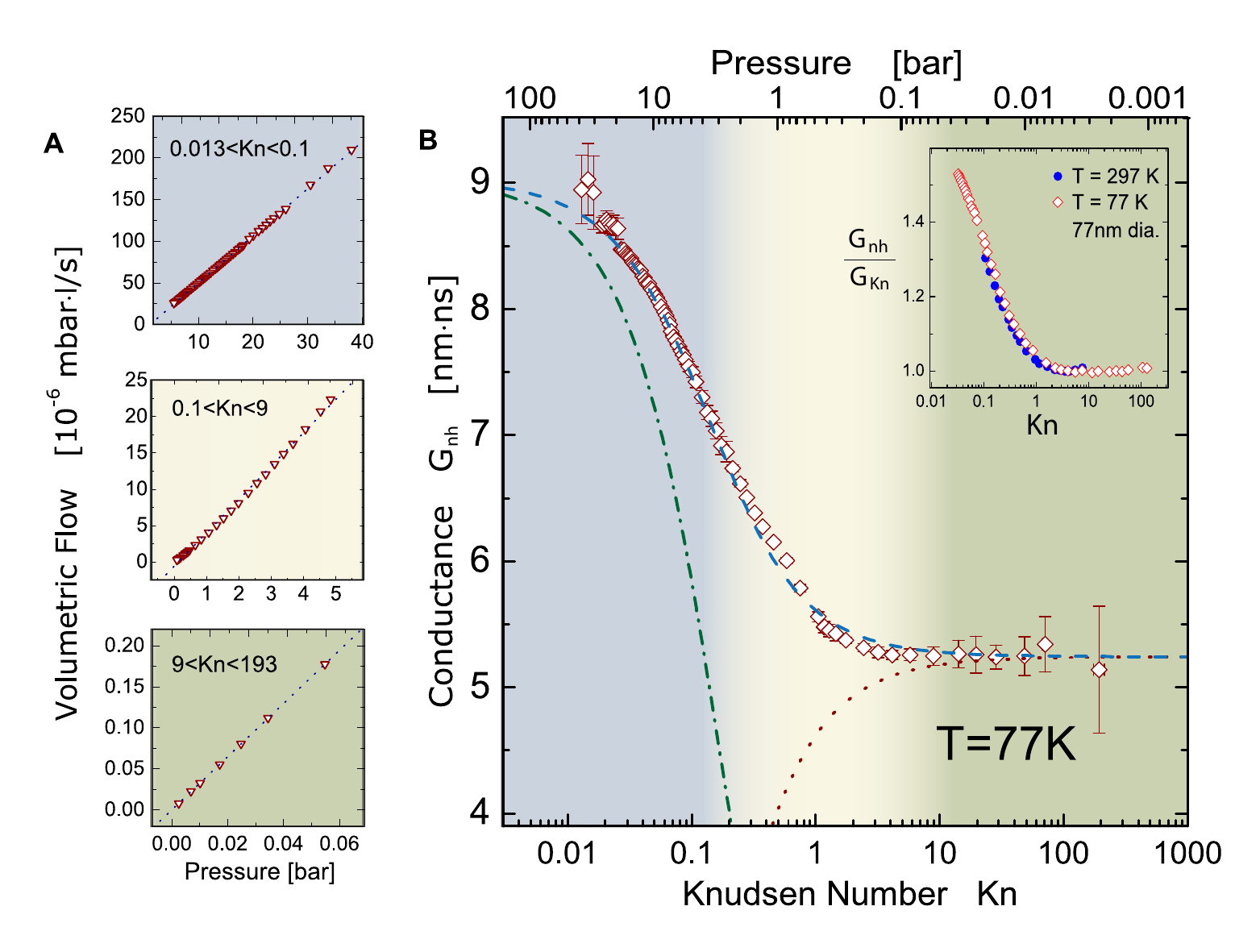}
\caption{ \textbf{A}, Volumetric flow (triangles) of helium gas versus the pressure $P_{S}$ measured by mass spectrometry 
at three different range of pressure (Knudsen number $Kn$). The dotted line is a linear fit to the low-pressure data. 
\textbf{B}, Conductance $G_{nh}$ (red diamonds) of the nanohole shown in Fig.1  versus the Knudsen number at 77K. The dotted red line is a fit of the data at high Knudsen number $Kn>10$ using a  free-molecular Knudsen flow model (with cutoff) and a nanopore radius of $50.0$ nm. The blue dashed  line is a fit using a unified flow model  (see text Eq. 1) for a finite-size pipe. The dot-dashed green line shows contribution to  the conductance from viscous collective flow. The inset shows the crossover transition for a 77 nm diameter nanohole at 297 K (blue data) and 77K (red data).   }\label{figure2}
\end{figure}


In a classical fluidic system, a dimensionless number can be used to sort out the relevant formalism to be applied for the calculation of flow across a channel.  This parameter is known as the  {\it Knudsen number} $Kn$  and is defined for a pinhole geometry as the ratio of the inter-particle mean free path $\lambda$  to the hole diameter $D$. At small Knudsen numbers $Kn \ll  1$, inter-particle collisions dominate, the transport is {\it collective} and  a formalism describing continuous media should be used. In the case of a viscous fluid,  the Navier-Stokes equations of fluid dynamics can be used to describe the flow. For $Kn \gg  1$, inter-particle collisional events are rare, the transport of particles is {\it individual} and a formalism based on the kinetic theory of statistical mechanics best describes the flow.  In the case of an ideal pinhole, the problem is very similar to that of a  photon escaping a black box:  transport of particles is given by a  Maxwell-Boltzmann speed distribution with each particle's  escape probability contributing to the current  flow out. In the context of molecular diffusion or fluidic transport, this mechanism is known as {\it effusion} for a hole, and {\it Knudsen diffusion} for a tube.\\

The mean free path in a gas can be calculated from kinetic theory with the relation  $ \lambda^{-1} = \sqrt{2} \pi d^2 n_v$ where $\pi d^2$ is the effective cross-section of collisions and $n_v$ is the volumic density determined from the pressure and temperature of the gas. In our experiment,  we use helium    which has $d = 2.2$x$10^{-10}$ m and a density calculated with the ideal gas law corrected with  known virial coefficients \cite{virial}. The mean free path can thus easily be tuned to be smaller or larger than the hole size.  In order to study the nonlinearity of the mass flow as the pressure is increased, we extract the  conductance $G_{nh}\equiv Q/\Delta P=Q/P_{S}$ from the measured flow.  The resulting conductances are plotted versus the Knudsen number extracted from $P_{S}$  in Fig. 2B (at $T=77$ K). Two distinct regimes can be seen in the data, one for $Kn>1$ for which $G_{nh}$  is  constant, and one for $Kn < 1$  where the conductance is continuously evolving. These data, spanning a measurement of $G_{nh}$ over four orders of magnitude in $Kn$, unambiguously show that a  crossover  transition in the mass flow occurs when the inter-particle mean free path approaches the nanohole diameter, as expected. For completeness, the inset of Fig.2B shows the crossover transition observed at $Kn\simeq 1$ in a 77 nm diameter nanohole.  Our results extend  those reported in an   {\it array} of silicon nanochannels \cite{GRUENER08} (with pore density $\sim 10^{11}$ $cm^{-2}$) to the single hole limit, revealing a clear picture of the dynamics of the flow in the crossover region.   \\


At low Kn, where the fluid is described by a continuum, the flow in a macroscopic capillary of radius $R$ and length $L$ can be modeled by the Navier-Stokes equations. For example, when $L\gg R$, one obtains the well-known pressure-dependent Poiseuille conductance of viscous flow, $G^{Poiseuille}_{viscous} = \frac{\pi \rho R^4}{8 \eta L}$, where $\rho=\rho(P,T)$ is the density of the fluid and $\eta=\eta(T)$ its dynamic viscosity,  which depends mostly on temperature. However, in a pinhole geometry where $R\sim L$, a model for short pipes must instead be used. Sherman has modeled the flow in short pipes \cite{SHERMAN92} by linearizing the Navier-Stokes equations, from which we can extract a conductance $G^{finite}_{viscous}=\frac{8 \pi \eta L}{\alpha } \left(\sqrt{1+\frac{\Delta P^2}{P_c^2}}-1\right) \frac{1}{\Delta P}$,
with  $P_{c}=\sqrt{\frac{64  \eta^2 L^2 k_B T}{\alpha m  R^4}}$  defining a critical pressure (and associated critical Knudsen number $Kn_{c}$) containing a geometry-dependent factor $\alpha$ accounting for the acceleration of the fluid at the boundary. As  a consequence of finite-size effects, we note that at large  $\Delta P$  the conductance reduces to $G_{viscous}^{\Delta P \rightarrow \infty} =  \pi R^{2}\sqrt{\frac{m}{\alpha k_{B} T}}$, which is pressure and viscosity independent. At high Kn, when the gas is very dilute and the mean free path is large compared to the capillary diameter, the gas is no longer forming a continuum  and the formalism of fluid mechanics should be replaced by a statistical approach. This leads to a Knudsen effusive flow given by $G^{finite}_{Knudsen} = \kappa \frac{3L}{8R} \left( \sqrt{\frac{32\pi m_{He}}{9k_BT}}\frac{R^3}{L} \right)$, where $\kappa$ is the Clausing factor which goes from 0 to 1 as the ratio $L/R$ goes from infinity to 0.  The term within parentheses is the Knudsen conductance for $L\gg R$. Building on the work of Ref. \cite{KARNIADAKIS05} for an infinitely long pipe, we write for a finite-size pinhole with $L/R\simeq 1$ a unified conductance describing the flow over the whole $Kn$ range as 

\begin{equation} G^{finite}_{unified} = G^{finite}_{viscous} + \frac{2}{\pi} arctan(\sigma\cdotp Kn) G^{finite}_{knudsen}
\end{equation} 
where an arctangent function was introduced as a cut-off  controlled by  the parameter $\sigma$. This {\it ad hoc} cut-off is necessary to account for the suppression of Knudsen diffusion  below $Kn  \simeq 0.1 $  where the onset of the dynamics in the continuum dominates over the transport of individual particles. \\

 Using the conductance data (above $Kn=10$)  for Knudsen flow (using a fixed pore length of 50 nm), we find a fitted radius (using the Knudsen conductance) of $50.0\pm 0.2$ nm, which is within 1\% of the radius determined by means of TEM microscopy.  Moreover, using the semi-phenomenological unified flow model given by Eq. 1, a fit of the conductance data (at $77$ K)  is shown in Fig. 2B with a dashed line and  with  fitted  parameters  $\alpha = 4.69\pm 0.06$ and $\sigma = 5.3\pm 0.1 $. The agreement with the data is excellent, and the individual contributions from Knudsen (with cut-off) and viscous flow to the conductance are shown with a dotted and dash-dotted line, respectively. This  good agreement between theory and data confirms the Knudsen theory predictions for a  $\sim 100$ nm nanohole both qualitatively (observation of pressure-independent conductance) and quantitatively (by direct extraction of the nanohole radius with the model).

\begin{figure}[tbp]
\includegraphics[width=0.86\linewidth,angle=0,clip]{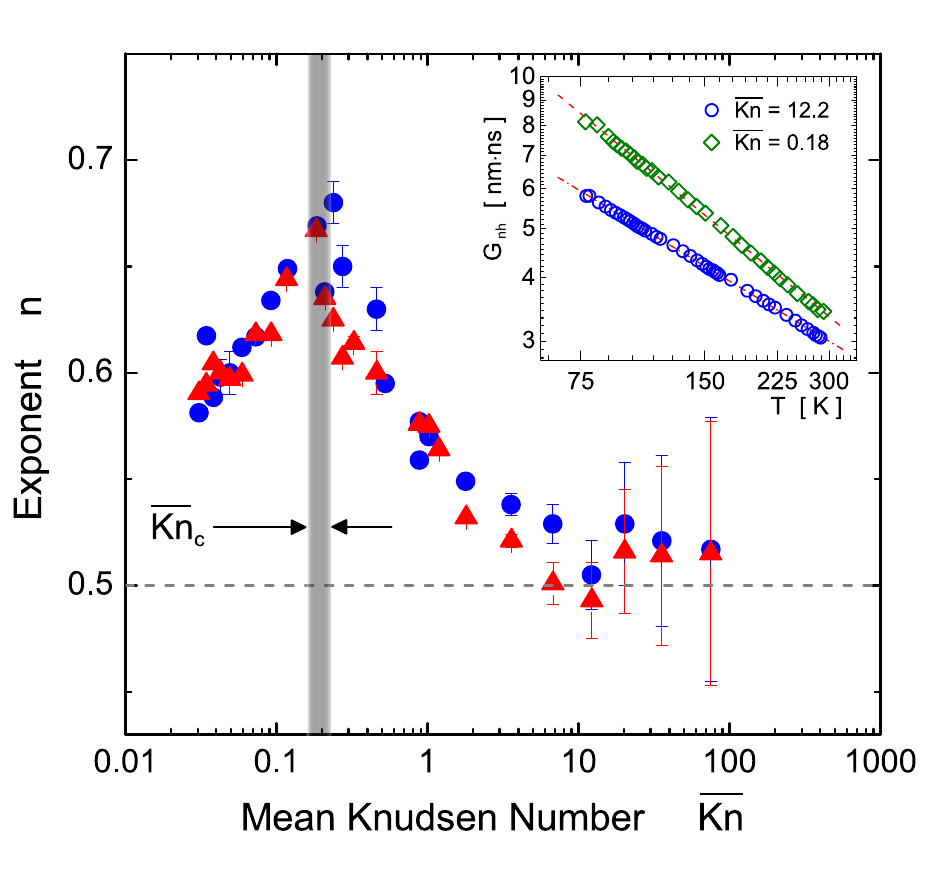}
\caption{Power-law exponent $n$ extracted from the conductance data taken upon warming (red triangles) and cooling (blue circles). The grey dashed line shows the inverse square root behavior expected for Knudsen free-molecular flow. The vertical grey bar indicates the critical Knudsen number $Kn_{c}$ 
(and its width the uncertainty) flagging the pressure at which the finite-size effects become predominant. The inset shows a log-log plot of the conductance versus the temperature at two Knudsen numbers and the dashed lines are fit giving $n$.}\label{figure3}
\end{figure}

In order to deepen our understanding of  fluidic transport in the transitional region, the temperature dependence of the conductance between $77$ K and $297$ K was studied in detail.  A fixed amount of gas was introduced in the cell and the temperature was slowly lowered to $77$ K over the course of several hours, and then warmed back up to room temperature while the mass flow was monitored. Upon cooling, both the volume and number of atoms were kept constant, so the pressure in the experimental cell was reducing according to the ideal gas law. For example, the top (squares) cooldown data shown in the  inset of Fig. 3 varied in pressure from $6.6$ bar at $297$ K to $6.08$ bar at $77$ K. This corresponds to Knudsen numbers decreasing by a factor of $\sim$3.4  from  0.286 to 0.083,  respectively, therefore averaging to $\overline{Kn} = 0.18$. For each cooling/warming procedure, we determine such a mean Knudsen number $\overline{Kn}$. The temperature of the gas during the procedure was inferred from a continuous measurement of the gas pressure, thereby forming a gas thermometer. The extracted conductances at initial pressures $P=6.6$ bar ($\overline{Kn} = 0.18$) and $P=97$ mbar ($\overline{Kn} = 12.2$) are shown on a log-log plot in the inset of Fig. 3  as green diamonds and blue circles, respectively. The linear behavior on the log-log scale allows us to write a power law of the form $G_{nh} \propto T^{-n}$, where $n$ is a power law exponent for the temperature dependence of the conductance. The power exponent $n$ extracted at each pressure and shown in the main panel of Fig. 3 is therefore plotted against  $\overline{Kn}$ for each cooling (circles) and warming (triangles) procedures.  Above $\overline{Kn} \simeq 10$, the power-law exponent $n$ recovers the square root behavior expected for Knudsen flow including finite-size effects, $G^{finite}_{Knudsen}\propto T^{-\frac{1}{2}}$. For $\overline{Kn}<10$,  a clear departure from an inverse square root is observed, with a peaking power $n\simeq 0.67$  at   $Kn \simeq 0.2$. At lower Knudsen number, a falling off of the exponent down  to $\sim 0.58$ at $Kn \simeq 0.03$ is seen, which corresponds to the highest differential pressure that the SiN  membrane could support prior to reaching the breaking point.

The fall-off trend below $Kn \simeq 0.2$ can easily be interpreted as the onset of finite-size conductance in the viscous regime.  
Interestingly, in the high-pressure  limit where $\Delta P\rightarrow  \infty$,  the conductance should asymptotically reach $G_{viscous}^{finite} \propto T^{-\frac{1}{2}}$ as a pure consequence of  finite-size effects. 
To support our interpretation, we have indicated in Fig. 3 by  a grey bar  (and its uncertainty by its width) the critical Knudsen number  $Kn_{c}$  obtained from  the $\alpha$ parameter fitted to the data  in Fig. 2 (and similar fit for data at 297K). From high-to-low Knudsen number, the rise and peak in the exponent $n$  in the transitional regime is consistent with the onset of  viscous flow towards the Poiseuille limit, up to $Kn_{c}$.  Beyond this point, the failure of the Poiseuille approximation  and the onset of finite-size effects forces the departure  from the Poiseuille regime and the fall-off of the exponent towards the finite-size, high-flow  limit.  We therefore conclude that  {\it i)} the expected temperature dependence for  Knudsen flow is  recovered above $\overline{Kn}$  $\simeq 10$, in complete agreement with the conductance data of Fig. 2  and {\it ii)}  the  crossover from an extended to a finite-size viscous flow in the conductance of a single nanohole is observed at $Kn$ in the vicinity of the critical Knudsen number, $Kn_{c}$.\\



 In conclusion, we have made a direct observation of the  smooth crossover between the collective transport of particles emanating from a continuum to a single-particle statistical flow occurring in a single nanohole.  Future experiments will focus on reducing the size of the nanofluidic channel even further,  towards the one-dimensional limit \cite{SATO05,LAMBERT08},  where quantum ballistic transport should give rise at low temperatures  to a conductance quantized in unit of $G_{m} = 2m^2/h$, the  quantum of conductance for mass flow.

This work has been supported by NSERC (Canada), FQRNT (Qu\'{e}bec),  and CIFAR. We thank J. Hedberg, R. Talbot, R. Gagnon and J. Smeros for technical assistance,  as well as I. Affleck, W. Mullin, M. Sutton, L. Pich\'e and H. Guo for illuminating discussions. We also thank the Center for the Physics of Materials at McGill (CPM) and the hospitality of the Centre for Characterization and Microscopy of Materials (CM$^{2}$) where the nanohole fabrication took place.

\end{document}